\documentclass[]{./aa}     
\usepackage{graphics}

\begin{document}
\thesaurus{07         
               07.09.1;  
               13.09.1;  
               12.12.1;  
               04.03.1;  
               04.19.1;  
               09.04.1;  
             }

\title {Calibration of the distance scale from galactic Cepheids:II}
\subtitle{Use of the HIPPARCOS calibration}
\titlerunning{Calibration of the distance scale  }

\author{Paturel G. \inst{1},
Teerikorpi P. \inst{2}, Theureau G. \inst{3}, Fouqu\'e P. \inst{4}, Musella I. \inst{5}, Terry J.N. \inst{1}}
\authorrunning{Paturel G., et al.}
\offprints{G. Paturel -
The compilation of raw data is available in electronic form
at CDS and on our anonymous ftp-server www-obs.univ-lyon1.fr (pub/base/CEPHEIDES.tar.gz).}

\institute{
              CRAL-Observatoire de Lyon,\\
              Avenue Charles-Andr\'e
              F69561 Saint-Genis Laval cedex, France \\
\and
              Turku University Observatory \\
              Tuorla     \\
              SF 21500 Piikkio, Finland \\
\and
              Laboratoire de Physique et de Chimie de l'Environnement \\
              3A Avenue de la Recherche scientifique   \\
              45071 Orleans cedex 02, France  \\
\and
              European Southern Observatory \\
              Casilla 19001      \\
              19 Santiago, Chile  \\
\and
              Osservatorio Astronomico di Capodimonte \\
              Via Moiariello 16, \\
              80131 Napoli, Italy \\
}

   \date{Received 8 August 2001 / Accepted 1 March 2002 }

   \maketitle

   \begin{abstract}
New estimates of the distances of 36 nearby galaxies is presented.
These are based on the calibration of the V- and I-band Period-Luminosity
relations for galactic Cepheids measured by the HIPPARCOS mission.
The distance moduli are obtained in a classical way.
The statistical bias due to the incompleteness of the
sample is corrected according to the precepts
introduced by Teerikorpi (1987).

We adopt a constant slope (the one obtained with LMC Cepheids).
The correction for incompleteness bias introduces an uncertainty
that depends on each galaxy. On average, this uncertainty is small
(0.04 mag) but it may reach 0.3 mag.
We show that the uncertainty due to the correction of the extinction
is small (propably less than 0.05 mag.).
The correlation between the metallicity and the morphological type
of the host galaxy suggests that we should reduce the application
to spiral galaxies in order to bypass the problem of metallicity.
We suspect that the adopted PL slopes are not valid for all morphological types
of galaxies. This may induce a mean systematic shift of 0.1 mag on
distance moduli.

A comparison with the distance moduli recently published by Freedman et al. (2001)
shows there is a {\bf reasonably} good agreement with our distance moduli.
   \end{abstract}
\section{Introduction}

We have started a new study of the kinematics
of the local universe (KLUN+) which aims at determining peculiar velocities for
nearby galaxies ($\approx 100/h$ Mpc). The radial component of such a peculiar velocity
is obtained by subtracting the Hubble flow from the observed radial velocity.
This means that the underlying Hubble flow must be determined free of
any sort of bias (systematic, distance
or direction dependent). The distances are obtained through the Tully-Fisher relation (1977)
by combining 21-cm line width measurements
(Nan\c{c}ay key-project) with infrared magnitudes (DENIS and 2MASS surveys).
The Tully-Fisher relation will be calibrated with some very
near galaxies ($< 25/h$ Mpc). This calibrating step is very important because
it will influence all forthcoming results. For this reason,
the distances of these calibrating galaxies must be determined carefully on the basis of
the Cepheid Period-Lumionosity relation (hereafter, PL relation) which remains the most
accurate method of stellar distance determination. Furthermore, the PL relation itself must
be calibrated from geometrical means, i.e. from galactic Cepheids.
In a previous paper (Paturel et al., 2002; paper I)
we obtained distances for 36 nearby galaxies by comparing, in a straightforward
way (the method of "sosie"), extragalactic Cepheids with galactic Cepheids
whose accurate distance moduli are available through the Barnes-Evans
method (Gieren, Fouqu\'e and Gomez, 1998; hereafter GFG).

Our present purpose is to calculate the distances through the classical PL
relation for the same galaxy sample using our calibration (Lanoix et al. 1999)
from the HIPPARCOS satellite (Perryman et al., 1997) which
measured geometrical parallaxes for a sample of nearby galactic Cepheids.
It has been shown (Pont et al., 1997; Lanoix et al. 1999) that
the treatment proposed by Feast and Catchpole (1997) to correct for
the Lutz and Kelker's bias (1973) gives an unbiased
calibration of the PL relation. 
The zero-point calibration is independent of the one used in paper I.


In section 2 the method of calculation is recalled and
applied to the V- and I-band measurements
described in paper I.  So,  distances are obtained for
1840 Cepheids belonging to 36 nearby galaxies.
In section 3 we discuss these results.

\section{Application to extragalactic Cepheids}
\subsection{The method of calculation}
The V-band magnitude $V$ can be corrected for extinction.
The corrected $V_o$ magnitude is given through the classical relation:
\begin{equation}
V_o= V  - R_V E_{(B-V)}
\end{equation}
where $R_V$ is the coefficient of the total to the differential
extinction (as tabulated e.g. by
Cardelli, Clayton \& Mathis, 1989 ;
Caldwell \& Coulson, 1987 ; Laney \& Stobie, 1994)
and $E_{B-V}$
is the $B-V$ color excess (difference between the observed and the
intrinsic $B-V$).

Similarly, the I-band magnitude can be corrected through the relation:
\begin{equation}
I_o= I  - R_I E_{(B-V)}
\end{equation}
where $R_I$ is the coefficient of the total to the differential
extinction for the I-band. Combining these two equations with the
Period-Luminosity-Color relation (PLC relation)
\begin{equation}
M_I = a logP +b +c (V-I)_o
\label{ami}
\end{equation}
and with the definition of the distance modulus
one obtains:

\begin{equation}
\label{fm}
\langle \mu \rangle = \frac{ \langle \mu_V \rangle - (R_V / R_I) \langle \mu_I \rangle }{ 1 - (R_V / R_I)}
\end{equation}
where,
\begin{equation}
\langle \mu_V\rangle  = \langle V \rangle  - (a_V \log P + b_V)
\end{equation}
\begin{equation}
\langle \mu_I \rangle = \langle I \rangle  - (a_I \log P + b_I)
\end{equation}

Let us recall that in these equations $\langle X \rangle $ means
{\it average over all the colors at the considered $\log P$}.
Because all these expressions are linear
it is equivalent to make the calculation for each individual Cepheid
and to deduce the mean $\langle \mu \rangle $ afterwards.
This is the method used in the present paper.
It is equivalent to the Wesenheit function method
as already emphasized by several authors (e.g., Tanvir 1997).

\subsection{The observational material}
In 1999 we constructed an Extragalactic Cepheid database
(Lanoix et al., 1999b).
The updated version contains 6685 measurements
for 2449 Cepheids in 46 galaxies.
The full contents of the extragalactic part is available in
electronic form as described in paper I.

Let us recall briefly the characteristics of the sample extracted from this database.
Each light curve has been inspected.
Only light curves classified as 'Normal' (see Lanoix et al., 1999b) are used.
Only the mean V and I band magnitudes are kept in the present study.
When several magnitudes are averaged from different sources, we keep the mean
only if the mean error is less than 0.05 magnitude.
The final sample results in 1840 extragalactic
Cepheids belonging to 36 galaxies. It is also available in electronic form
(see paper I).
The source codes of measurements are given for each galaxy in Table \ref{cati}. 
The full references appear in the bibliography with their codes.

\begin{table}
\caption{Sample of extragalactic Cepheids.
{\bf Column 1:} Name of the host galaxy;
{\bf Column 2:} list of reference codes.}
\begin{tabular}{ll}
\hline
galaxy &Cepheid Reference code\\
\hline
IC1613   & Fr88a Sa88a  \\
IC4182   & Sah94 Gib99 \\
LMCogle  & Uda99  \\
NGC1326A & Pro99 \\
NGC1365  & Sil98 \\
NGC1425  & Mou99 \\
NGC2090  & Phe98 \\
NGC224   & Fre90 \\
NGC2541  & Fer98 \\
NGC300   & Fre92 Wal88 \\
NGC3031  & Fre94 \\
NGC3109  & Mus98 Sa88b \\
NGC3198  & Kel99 \\
NGC3319  & Sak99 \\
NGC3351  & Gra97 \\
NGC3368  & Tan95 Gib99 \\
NGC3621  & Raw97 \\
NGC3627  & Sah99 Gib99 \\
NGC4258  & Mao99 \\
NGC4321  & Fer96 \\
NGC4414  & Tur98 \\
NGC4496A & Sh96c Gib99 \\
NGC4535  & Mac99 \\
NGC4536  & Sh96a Gib99 \\
NGC4548  & Gra99 \\
NGC4603  & New99 \\
NGC4639  & Sah97 Gib99 \\
NGC4725  & Gib98 \\
NGC5253  & Sah95 Gib99 \\
NGC5457  & Alv95 Kel96 \\
NGC598   & Chr87 Fre91 Kin87 Sa88a \\
NGC6822  & Gal96 Kay67 \\
NGC7331  & Hug98 \\
NGC925   & Sil96 \\
SEXA     & Pio94 \\
SEXB     & Pio94 Sa85b \\
\hline
\end{tabular}
\label{cati}
\end{table}

\subsection{Adopted PL relations}
From HIPPARCOS measurements of 238 galactic Cepheids
 \footnote{174 Cepheids in I-band.}
we obtained unbiased V- and I-band
Period-Luminosity relations (Lanoix et al. 1999) using the
treatment described by Feast and Catchpole (1997):
\begin{equation}
\langle M_V \rangle = -2.77 \log P - 1.44 \pm 0.05
\label{PLV}
\end{equation}
\begin{equation}
\langle M_I \rangle = -3.05 \log P - 1.81 \pm 0.09.
\label{PLI}
\end{equation}
Let us recall that the slopes of the PL relations could not be
determined from HIPPARCOS measurements. Only the zero points 
(i.e. the mean absolute magnitude at a mean Period) have been fixed.
The V-band slope $a_V=-2.77$ was adopted from a mean
of different values obtained
for LMC (Caldwell \& Laney 1991; Madore \& Freedman 1991; Tanvir 1997; Gieren,
Fouqu\'e \& Gomez 1998). The I-band slope  $a_I=-3.05$ resulted from $a_V$ and 
from the slope ($0.28$) of the mean Period-Color relationship.
Eq. \ref{fm} is applied to the 1840 Cepheids of our sample.
We adopt the ratio ${R_V}/{R_I}=1.69$ (Cardelli et al., 1989). For each extragalactic
Cepheid of each host galaxy we plot the apparent distance modulus {\it vs.}
$\log P$.

It is important to emphasize that we adopt the LMC slopes and assume 
that it is universal and bias free.
The question of the choice of the slope will be addressed separately.

\subsection{The correction for incompleteness bias}
Sandage (1988) noticed that truncating a complete sample of Cepheids in LMC
changes the slope of the resulting PL relation. After Sandage this question remained untouched
for several years. Kelson (1996) mentioned the incompleteness bias 
and suggested one use the inverse slope to correct the effect.
Then, the effects of the bias were described from observation
(Paturel et al., 1997a) and from simulation (Lanoix et al., 1999a).
This effect is not negligible, e.g., it can affect the distance modulus by 0.4 magnitude for
a galaxy like NGC4536. 
One way to reduce the effect consists in using a magnitude limiting cut-off (Freedman et al., 2001).
Another way consists of fitting
the bias (the variation of the distance modulus with $\log P$). 
A biased distance modulus appears smaller by the quantity
(Teerikorpi, 1987 ; see paper I):

\begin{equation}
\Delta \mu=  - \sigma {\sqrt{\frac{2}{\pi}} \frac {e^{-A^2}}{1+erf(A)}}
\label{deltamu}
\end{equation}

where

\begin{equation}
A= \frac {V_{lim} -\mu - a_v \log P - b_v} {\sigma \sqrt{2}}
\end{equation}

and

\begin{equation}
erf(x) = \frac{2}{\sqrt{\pi}} \int_0^x e^{-t^2} dt.
\end{equation}

Here we assume, consistent with Lanoix et al. (1999), paper I and Freedman et al. (2001), 
that the relevant dispersion $\sigma$ is the scatter in the
mean absolute magnitude at a fixed $P$.
An iterative process is used because the distance modulus
must be known to calculate the bias. We start with an initial
distance modulus estimated from a simple mean
and then, by scanning distance
modulus within $\pm 1$ mag. around the initial value, we search for
the best fit.
For each galaxy, the limiting magnitude $V_{lim}$ is taken from paper I;
The initial value of the dispersion ($\sigma$) is calculated
from the uncorrected distance moduli;
Then it is adjusted at each step of the iterative process.

\begin{figure*}[p]
\resizebox{\hsize}{!}{\includegraphics*{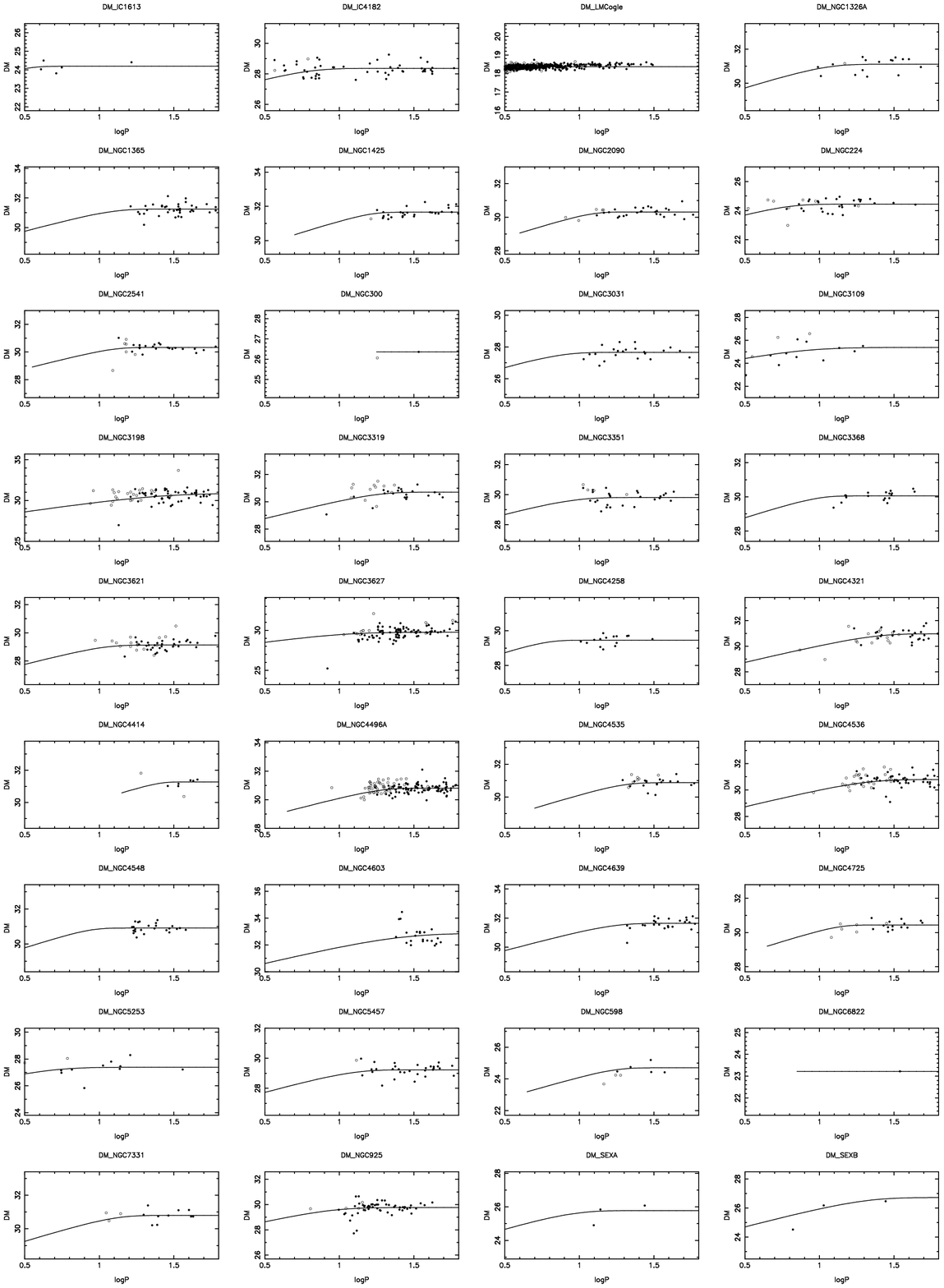}}
\caption{Distance moduli (y-axis) from the PL relations {\it vs.}
$\log P$ (x-axis) for each host galaxy.
}
\label{fig001}
\end{figure*}

For each of the 36 host galaxies we plot the apparent distance moduli
given through Eq. \ref{fm} as a function of $\log P$ and we superimposed
the bias curve obtained after the last iteration.
This result appears in Figure \ref{fig001} and Table \ref{fin}.
The standard error on the distance moduli is an internal error.

We want now to see how the results are modified when different PL relations
are used.  We want also to test the stability of the solution.

\section{Discussion}
\subsection{Comparison with an independent treatment.}
Freedman et al. (2001) recently published their
HST key-project distance moduli (The Hubble Space Telescope Key Project
on the Extragalactic Distance Scale)
calibrated with LMC, assuming $\mu(LMC) =18.5$.
The agreement between HSTKP distance moduli and those calculated here
is good for the 31 galaxies in common.
We do not confirm the tendency found in paper I that their distance moduli
are smaller than ours above $\mu=30$.
A direct regression  of our distance moduli {\it vs.} HSTKP distance moduli $\mu(HSTKP)$ 
(uncorrected for metallicity effect, denoted $\mu_o$ and presented in column 8 of 
Table 4 in the paper by Freedman et al.) 
leads to a slope which is not significantly different from one
($1.012 \pm 0.005$) and a zero point difference which is not significantly
different from zero ($-0.027 \pm 0.016$) at the 0.01 probability level (the
Student's t-test requires $t_{0.01}(\nu=30) > 2.75$). 
The standard deviation is $\sigma=0.090$. Thus,
assuming both determinations carry the same uncertainty, this means that our
distances are good within $0.09/ \sqrt{2}=0.06$ magnitude. This is not perfectly
exact because both solutions are not fully independent (except for the
zero-point calibration). In particular, we use the same ratio $R_V / R_I = 1.69$
(Cardelli , Clayton \& Mathis, 1989) and most of the observations are the same.

\subsection{Influence of the choice of the PL relations.}
\begin{figure}[!]
\resizebox{\hsize}{!}{\includegraphics*{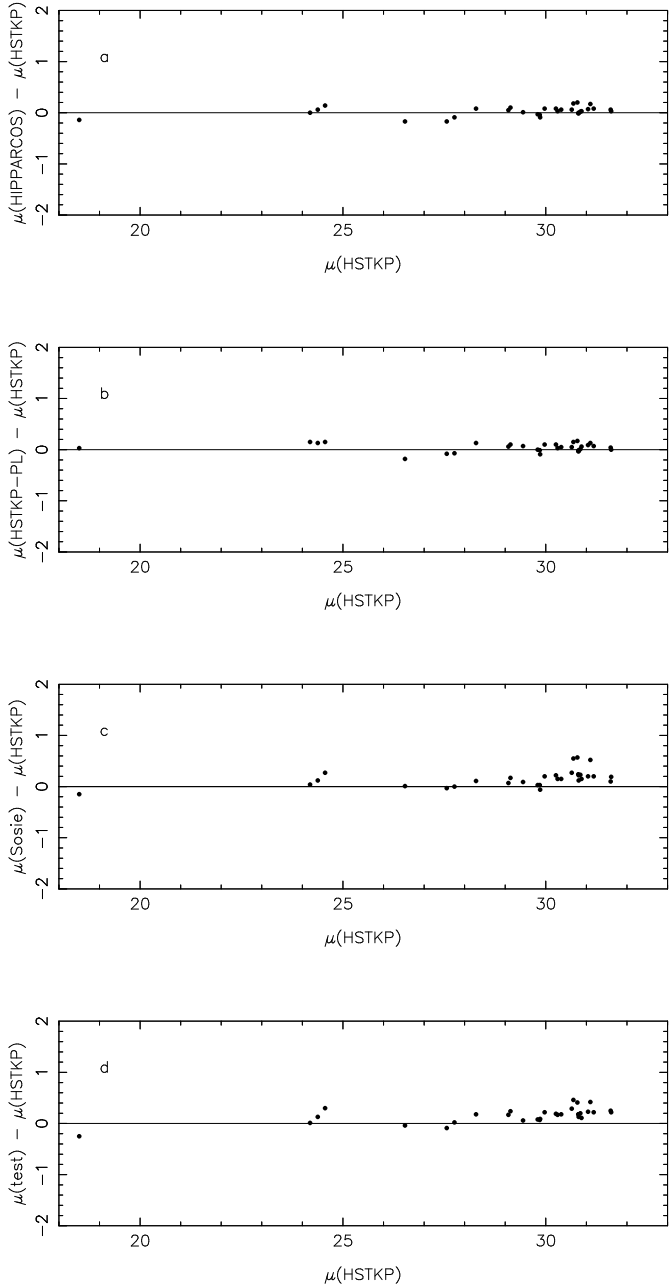}}
\caption{Comparison between the distance moduli from
different solutions (Table \ref{comp}) and the HSTKP distance moduli
$\mu_o$. The different solutions are applied on the same data.
{\bf a)}The first solution (HIPPARCOS calibration, this paper) is in reasonnably 
good agreement with the HSTKP solution.
{\bf b)} The second solution (HSTKP-PL relations applied on the present data). 
{\bf c)}  The third solution (SOSIE method applied on the same data, paper I) shows a 
departure from the HSTKP solution, especially at large distance ($\mu > 30$).
{\bf d)} The fourth solution (test solution with larger PL slopes) shows the same 
trend as the third one and suggests
that the choice of the PL slopes may have an important effect, especially at 
large distances. 
}
\label{fig003}
\end{figure}

The agreement between our solution and the HSTKP one may appear strange
because our Relations \ref{PLV} and \ref{PLI} do not differ very much from the old
calibration
{\footnote{The old Madore and Freedman (1991) PL relations were:
\begin{equation}
\langle M_V \rangle = -2.76 \log P - 1.40
\end{equation}
\begin{equation}
\langle M_I \rangle = -3.06 \log P - 1.81
\end{equation}
}}
(Madore and Freedman, 1991) which was revised
(Freedman et al., 2001) using Udalski et al. (1999) results on LMC.
The new HSTKP calibration is :
\begin{equation}
\label{FPLV}
\langle M_V \rangle = -2.76 \log P - 1.45
\end{equation}
\begin{equation}
\label{FPLI}
\langle M_I \rangle = -2.96 \log P - 1.94
\end{equation}
Thus, it will be interesting to apply these new relations to our own sample to see
if the agreement is not fortuitously due to the data.
We will keep $R_V /R_I= 1.69$. This is
the value adopted in the HSTKP, in our paper I and in the present paper .
In Table \ref{comp} we give the mean shift
$\langle \Delta \mu \rangle = \langle \mu \rangle -  \langle \mu(HSTKP)\rangle $
for different solutions applied to our data.
The result is also shown in Figs. \ref{fig003}a-d
The definitions of the different solutions are the following:

{\bf Solution \#1: HIPPARCOS : }The PL relations  are Rel. \ref{PLV} and \ref{PLI}.

{\bf Solution \#2: HSTKP-PL } : The PL relations  are those adopted
by Freedman et al. (2001), i.e., Rel. \ref{FPLV} and \ref{FPLI}.

{\bf Solution \#3: GFG-SOSIE } : This is the solution from paper I.
Slopes and zero-points are not required
explicitly. The calibration is based on the GFG sample.

{\bf Solution \#4: test  } : This is a test of a change of slopes (the zero-points being
recalculated from the HIPPARCOS calibration as described in footnote 3). This solution is
discussed below.

\vspace{0.5cm}
The distance moduli found with the HSTKP-PL relations are similar
to the final ones  published by the HSTKP team (see Figure \ref{fig003}b) but
the difference $\langle \Delta \mu \rangle = 0.045 \pm 0.015$ is significant
at the $3$-$\sigma$ level.
This difference  can be explained by the fact
that the data are not exactly the same and that the correction for the
incompleteness bias is made in a different way. 
Part of the difference can be explained as a consequence of the fact that changes
in the photometric zero point adopted by the HSTKP (Stetson 1998) have not been reflected
in the Lanoix et al. compilation which is used in this paper. The Lanoix
compilation uses slightly different zero points for different galaxies. 
For 50\% of the galaxies of the present sample, the distance moduli are 0.06 mag larger than 
the distance moduli used by Freedman et al. 2001.
The mean observed departure ($0.045$), although significant, is relatively small in comparison 
with the departure that could be due to an uncertainty on the PL slope, 
as illustrated by Figs. \ref{fig003}a-d and Table \ref{comp}.

The LMC distance
modulus is retrieved at $\mu (LMC)= 18.5$ as assumed for the HSTKP calibration.
In fact the difference between the adopted PL relations of solutions 1 and 2
is smaller than it appears. Indeed,
if one forces a slope of $-2.76$ in V-band (respectively, $-2.96$ in I-band)
on the HIPPARCOS zero-point which was obtained at a mean $\log P = 0.82$
(see Lanoix et al., 1999) one obtains the corrected PL relations:
\footnote{With obvious notation the
new zero-point is\\
 $b' = b + (a-a') \langle \log P \rangle$}.
\begin{equation}
\langle M_V \rangle = -2.76 \log P - 1.45
\end{equation}
\begin{equation}
\langle M_I \rangle = -2.96 \log P - 1.88
\end{equation}
These corrected PL relations do not differ very much from the new HSTKP PL ones.
This explains the relatively good agreement between solution 1 and the HSTKP
results (or, equivalently, with Solution \#2).

On the contrary, the last solution (GFG-SOSIE) shows a departure from the first two
solutions (HIPPARCOS and HSTKP-PL) especially above 10 Mpc ($\mu =30$).
It seems that one retrieves the dilemma emphasized in paper I that either the
HSTKP distance moduli may have a small residual bias or that the
GFG sample may overestimate the absolute magnitude for long periods.

If the PL slopes are changed into $-3.0$ for the V-band
(respectively, $-3.3$ for the I-band) as suggested by the results
of GFG or Laney and Stobie (1994), the zero-points being still recalculated from
our HIPPARCOS calibration, then the results (Solution \#4 in Table \ref{comp})
are compatible with those of
our Solution \#3 (the mean shift $\Delta \mu= +0.163$ while
the Solution \#3 gives $\Delta \mu= +0.161$).
Thus, we suspect that the PL slopes adopted from LMC are not valid for all kinds of
galaxies.
This question will be discussed elsewhere. Here, we will adopt our Solution \#1.
The uncertainty due to the slope will be discussed in the error budget.

\begin{table*}
\caption{Distance moduli calculated from different solutions (see text)
with $R_V / R_I = 1.69$.}
\begin{tabular}{llllll}
\hline
Solution     & $a_V$ & $b_V$ & $a_I$ & $b_I$  & $\langle \Delta \mu \rangle = \langle \mu - \mu(HSTKP)\rangle $  \\
\hline
\#1: HIPPARCOS     &$-2.77$ &$-1.44$ &$-3.05$ &$-1.81$  & $0.027 \pm 0.016$\\
\#2: HSTKP-PL     &$-2.760$&$-1.458$&$-2.962$&$-1.942$ & $0.045 \pm 0.015$\\
\#3: GFG-SOSIE$^a$        &\multicolumn{4}{l} {slopes and ZP not required}  & $0.161 \pm 0.029$ \\
\#4: test         &$-3.0$&$-1.02$ &$-3.3 $&$-1.44$ & $0.163 \pm 0.026$\\
\hline
\multicolumn{6}{l}{a) based on the GFG sample}\\
\end{tabular}
\label{comp}
\end{table*}

\subsection{Influence of the incompleteness bias}
As we explained above, the determination of $V_{lim}$ may affect the
correction of the incompleteness bias.
In order to evaluate the mean effect we repeated the calculation of distance
moduli varying  $V_{lim}$ over the range $(V_{lim}-0.5, V_{lim}+0.5)$.
The mean changes of distance moduli, $\Delta \mu ^-$ and $\Delta \mu^+$ respectively,
are given in Table \ref{test}. The mean change is less than $0.05$ magnitude
when $V_{lim}$ changes by $0.5$ magnitude. Nevertheless, individual changes
may be larger than this mean value. Hence, for each galaxy we give the
individual  $\Delta \mu ^-$ and $\Delta \mu^+$ in Table \ref{fin}.
The change is generally smaller than $0.1$ magnitude. Because it depends
on each individual galaxy, it introduces a random error. 
Assuming that the uncertainty on the limiting magnitude is $\pm 0.5$, the resulting 
error will be calculated for each galaxy from the relation
\footnote{For a Gaussian distribution $\mathcal{G}(\overline{x}, \sigma)$, 
the standard deviation is 
$\sigma= 1.25 \langle | \Delta x | \rangle $.}
$\sigma_{bias}= 1.25 | \Delta \mu |$.
For $\Delta \mu $  we will adopt the maximum between $\Delta \mu^+$ and $\Delta \mu ^-$.
It will be taken into account in the error budget.

\subsection{Influence of the ratio $R_V/R_I$}
In order to check the stability of our adopted solution,
we repeated the calculations with a variation of the ratio
${R_V}/{R_I}$ by $\pm$ 0.2 (about 10\%).  The results are summarized
in Table~\ref{test}, where we give the difference between the
mean distance moduli obtained with different ${R_V}/{R_I}$ ratios
and our reference solution based on ${R_V}/{R_I}=1.69$.
A change of ${R_V}/{R_I}$ by $\pm$10\% changes the distance moduli
by less than 0.1 magnitude. If the ${R_V}/{R_I}$ is not the same
for all galaxies, this introduces a dispersion of the calculated
distance moduli, but not a systematic effect.
Assuming that the uncertainty on  ${R_V}/{R_I}$ is $\pm 0.2$, 
the resulting error is about $\sigma_{extinction}= 1.25 |\Delta \mu | \approx 0.09$.
It will be taken into account in the error budget.

\begin{table}
\caption{Test of the stability of the results. We give the departure
from our reference solution for several ${R_V}/{R_I}$ ratios.}
\begin{tabular}{rrr}
\hline
$\Delta V_{lim}$&${R_V}/{R_I}$  &    $\mu - \mu_{ref}$  \\
\hline
0.0  &1.89           &    $-0.05 \pm 0.04$ \\
0.0  &1.79           &    $-0.03 \pm 0.01$ \\
0.0  &1.69           &    $0$              \\
0.0  &1.59           &    $+0.04 \pm 0.03$ \\
0.0  &1.49           &    $+0.09 \pm 0.06$ \\
\hline
$-0.50$  &1.69           &    $-0.04 \pm 0.14 $              \\
$-0.25$  &1.69           &    $-0.02 \pm 0.09 $              \\
$+0.25$  &1.69           &    $+0.02 \pm 0.04 $              \\
$+0.50$  &1.69           &    $+0.03 \pm 0.07 $              \\
\hline
\end{tabular}
\label{test}
\end{table}

\subsection{Influence of metallicity}
The problem of metallicity was first recognized by Iben (1967).
According to Freedman and Madore (1990) the coefficients of
the PL relation are slightly dependent on metallicity. Thus, the zero-point
of the extragalactic distance scale would be slightly dependent on the
metallicity. However, it has been argued that
the correction of interstellar absorption is particularly
sensitive to the metallicity (Beaulieu et al., 1997).

Most empirical investigations (Gould et al. 1994; Sasselov et al., 1997; 
Kennicut et  al., 1998) ) find a positive effect ranging from 0.24 
(Kennicut et al., 1998) to 0.56 (Gould et al., 1994).
More recently, Udalski (1999) confirmed Freedman and Madore (1990) result
that the metallicity effect is negligible.
In the theoretical approaches, the results of linear computations 
(Chiosi et al. 1993; Sandage et al. 1999; Alibert et al. 1999) suggest a 
small negative effect. However, using non-linear models (Bono et al., 2000)
Caputo et al. (2000) found a positive effect.

\begin{figure}[!]
\resizebox{\hsize}{!}{\includegraphics*{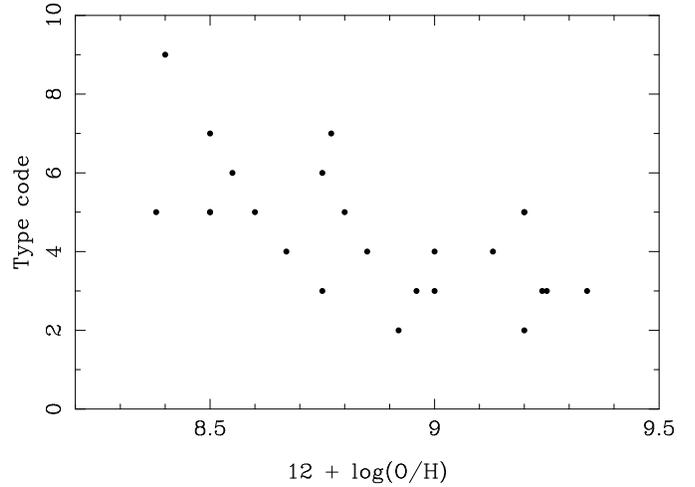}}
\caption{Correlation between the morphological type code
of the host galaxy and its metallicity $12 + \log O/H$.
}
\label{fig005}
\end{figure}

Owing to these puzzling results, we do not expect to solve the problem in the classical way.
Instead, we will avoid it by considering that
the method is valid only for galaxies with nearly the same metallicity
as the calibrating Cepheids (i.e., nearly Solar metallicity).
Indeed, there is a clear correlation (Fig. \ref{fig005}) between the morphological
type code of the host galaxy and its metallicity $12 + \log O/H$ as listed by
Caputo et al. (2000). Thus, we will consider that only spiral galaxies
over the range Sa-Scd (i.e., type codes 1-7)  should be considered as reliable. 
When this restriction is applied, we may consider that the uncertainty due to 
metallicity is negligibly small.

\subsection{Error budget}
In Table \ref{fin} we summarize our determinations of distance moduli
for 36 galaxies calibrated with galactic Cepheids.
The provisional distance moduli are calculated from a weighted mean of
our two determinations (GFG-SOSIE and HIPPARCOS).
The weight is  the inverse of the square of the individual mean error.
The final error on the mean distance modulus is the 'actual error'
(Paturel et al., 1997)  which takes into account the
individual errors (uncertainty due to the data) and the discrepancy between the solutions
(uncertainty due to the adopted zero-point). This uncertainty will be designated 
$\sigma_{zero-point}$.
The uncertainties due to the incompleteness bias correction and to the extinction correction
are added.
The uncertainty resulting from a possible metallicity effect will be neglected
but galaxies with a morphological type out of the accepted
range (Sa-Scd) are given in parenthesis.

We believe that another source of uncertainty can result from the choice of the
slope. This is partially taken into account in $ \sigma_{zero-point}$ because
the GFG-SOSIE solution does not require knowledge of the slope. This additional
uncertainty is $ 1.25 |\Delta \mu | / \sqrt{2}$ (i.e. $\approx 0.1$ mag  with 
$\Delta \mu \approx 0.11 $).

Finally, the total uncertainty  $\sigma_{total}$ (internal plus external) is calculated for each individual
galaxy from:

\begin{equation}
\sigma_{total}^2 = \sigma_{zero-point}^2 + \sigma_{bias}^2 + \sigma_{extinction}^2 + \sigma_{slope}^2
\end{equation}
This estimate is given with the provisional distance modulus in Table \ref{fin}.

\begin{figure}[!]
\resizebox{\hsize}{!}{\includegraphics*{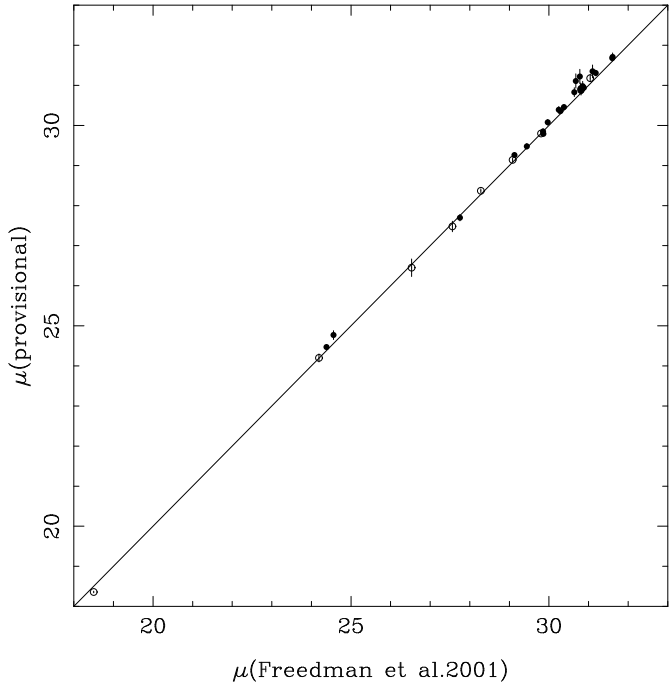}}
\caption{Comparison of the distance moduli from
Freedman et al., 2001 and from this paper.
The general agreement is satisfactory despite the being calibration
is independent. Late type galaxies are represented with
open circles.
}
\label{fig002}
\end{figure}

A comparison with the HSTKP distance moduli is presented in
Fig. \ref{fig002}. Excepting two galaxies at large distances
(NGC4321 and NGC3198 noted with a $(:)$ in Table \ref{fin}),
the agreement is good.

\section{Conclusion}
The preliminary  distance moduli obtained in the first two papers
of this series were analyzed to search for possible residual bias.
Distance moduli from paper I and from this paper agree reasonably well
within 0.1 magnitude, although they are
based on two independent calibration (GFG-SOSIE and HIPPARCOS) and two
independent methods (Sosie and classical PL relations).

The discussion of the stability of the solution shows that the slope
of the PL relations is still under question. The LMC slope in V-band
(and maybe for all late type galaxies) seems well fixed
($a_V \approx -2.77$) but several studies (GFG, LS),
including this paper, show that the slope for galactic Cepheids (and
maybe for all Sa-Scd galaxies) could be steeper ($a_V \approx -3$).
If one adopts a slope $a_V = -2.76$ the results are in good
agreement with the results of HSTKP. If one adopts a slope $a_V = -3.0$
the distance moduli must be increased, on average, by 0.1 mag.

When all sources of errors are taken into account, the mean
standard deviation of the final distance modulus is about 0.20 magnitude. 

The correlation between metallicity and morphological type of
hosts galaxies suggests to limit the validity of our distances to
spiral galaxies (Sa-Scd) that have the same metallicity as our
calibration sample.

For NGC4258 our distance modulus, $\mu = 29.48 \pm 0.16$, is
compatible with the maser determination $\mu = 29.28 \pm 0.15$
(Herrnstein et al., 1999) and it is in good
agreement with the revised distance modulus $\mu = 29.48 \pm 0.15$
(Newman et al., 2001).

If it is confirmed that the slopes of PL relations have to be adapted to
the morphological type of each host galaxy, these
distance moduli could be modified. 
Further, 
using the local Hubble flow for providing independent reference distances,
an additional analysis of the bias on primary calibration (Teerikorpi\& Paturel, 2002)
suggests the existence of another bias. The problem of distance calibration 
is not yet resolved.

\begin{table*}
\caption{Distance moduli.
{\bf Column 1:} Name of the host galaxy;
{\bf Column 2:} Morphological type code (de Vaucouleurs et al., 1991).
{\bf Columns 3:} Distance modulus from paper I (calibration with galactic Cepheids
from GFG sample) with its internal error.
{\bf Column 4:} The distance modulus from the present paper (calibration with galactic Cepheids
from HIPPARCOS) with its internal error.
{\bf Columns 5-6:} changes of distance modulus when the limiting 
magnitude is changed by $\pm 0.5$ mag.
{\bf Column 7:} The distance modulus from HSTKP with its random uncertainty
(columns 8 and 9 in Freedman et al. 2001).
{\bf Column 8:} Adopted distance modulus with its total error.
Distance moduli given in parenthesis correspond to late type galaxies.
}
\begin{tabular}{lrrrrrrr}
\hline
galaxy  & Type & GFG-SOSIE    &   HIPPARCOS   & $\Delta \mu^-$&$\Delta \mu^+$       & HSTKP   &  Provisional $\mu$      \\
\hline
IC1613  & 10 &$ 24.23\pm  0.19$ &   24.19$\pm$    0.11	 &   0.09 &  -0.02	&   24.19$\pm$0.15 & ($24.20 \pm  0.20$)  \\
IC4182  &  9 &$ 28.39\pm  0.07$ &   28.36$\pm$    0.06	 &  -0.02 &  -0.03	&   28.28$\pm$0.06 & ($28.37 \pm  0.15$)  \\
LMCogle &  9 &$ 18.35\pm  0.03$ &   18.37$\pm$    0.00	 &   0.01 &   0.00	&   18.50          & ($18.37 \pm  0.14$) \\
NGC1326A&  9 &$ 31.24\pm  0.09$ &   31.11$\pm$    0.10	 &   0.05 &  -0.04	&   31.04$\pm$0.10 & ($31.18 \pm  0.17$)  \\
NGC1365 &  3 &$ 31.38\pm  0.07$ &   31.26$\pm$    0.06	 &   0.00 &  -0.01	&   31.18$\pm$0.05 & $31.31 \pm  0.15$  \\
NGC1425 &  3 &$ 31.70\pm  0.06$ &   31.66$\pm$    0.05	 &   0.07 &  -0.05	&   31.60$\pm$0.05 & $31.68 \pm  0.17$  \\
NGC2090 &  5 &$ 30.44\pm  0.07$ &   30.32$\pm$    0.05	 &   0.02 &   0.00	&   30.29$\pm$0.04 & $30.36 \pm  0.15$  \\
NGC224  &  3 &$ 24.50\pm  0.08$ &   24.44$\pm$    0.07	 &   0.14 &  -0.04	&   24.38$\pm$0.05 & $24.47 \pm  0.23$  \\
NGC2541 &  5 &$ 30.47\pm  0.07$ &   30.33$\pm$    0.06	 &  -0.02 &  -0.03	&   30.25$\pm$0.05 & $30.39 \pm  0.16$  \\
NGC300  &  6 &$ 26.54\pm  0.29$ &   26.36$\pm$    0.29	 &   0.00 &  -0.15	&   26.53$\pm$0.07 & $26.45 \pm  0.32$  \\
NGC3031 &  2 &$ 27.75\pm  0.10$ &   27.66$\pm$    0.08	 &   0.03 &  -0.02	&   27.75$\pm$0.08 & $27.70 \pm  0.16$  \\
NGC3109 &  9 &$ 25.10\pm  0.16$ &   25.38$\pm$    0.22	 &  -0.08 &   0.16	&         & ($25.20 \pm  0.31$)  \\
NGC3198 &  5 &$ 31.23\pm  0.07$ &   30.86$\pm$    0.10	 &   0.05 &  -0.02	&   30.68$\pm$0.08 & $31.11:\pm  0.23$  \\
NGC3319 &  5 &$ 30.91\pm  0.06$ &   30.70$\pm$    0.08	 &   0.03 &   0.02	&   30.64$\pm$0.09 & $30.83 \pm  0.18$  \\
NGC3351 &  3 &$ 29.88\pm  0.08$ &   29.81$\pm$    0.09	 &   0.04 &   0.04	&   29.85$\pm$0.09 & $29.85 \pm  0.16$  \\
NGC3368 &  2 &$ 30.17\pm  0.10$ &   30.05$\pm$    0.06	 &   0.06 &  -0.02	&   29.97$\pm$0.06 & $30.08 \pm  0.17$  \\
NGC3621 &  6 &$ 29.15\pm  0.07$ &   29.13$\pm$    0.06	 &   0.02 &   0.03	&   29.08$\pm$0.06 & $29.14 \pm  0.15$  \\
NGC3627 &  3 &$ 29.80\pm  0.06$ &   29.77$\pm$    0.07	 &   0.03 &  -0.02	&   29.86$\pm$0.08 & $29.79 \pm  0.15$  \\
NGC4258 &  4 &$ 29.53\pm  0.10$ &   29.45$\pm$    0.07	 &   0.05 &   0.00	&   29.44$\pm$0.07 & $29.48 \pm  0.16$  \\
NGC4321 &  4 &$ 31.35\pm  0.06$ &   30.98$\pm$    0.08	 &   0.00 &  -0.06	&   30.78$\pm$0.07 & $31.22:\pm  0.24$  \\
NGC4414 &  5 &$ 31.62\pm  0.09$ &   31.27$\pm$    0.05	 &   0.13 &  -0.06	&   31.10$\pm$0.05 & $31.35 \pm  0.26$  \\
NGC4496A&  7 &$ 31.03\pm  0.04$ &   30.81$\pm$    0.03	 &   0.01 &   0.03	&   30.81$\pm$0.03 & $30.89 \pm  0.18$  \\
NGC4535 &  5 &$ 31.08\pm  0.07$ &   30.87$\pm$    0.07	 &  -0.07 &   0.03	&   30.85$\pm$0.05 & $30.98 \pm  0.20$  \\
NGC4536 &  4 &$ 31.04\pm  0.06$ &   30.79$\pm$    0.07	 &  -0.02 &   0.00	&   30.80$\pm$0.04 & $30.93 \pm  0.19$  \\
NGC4548 &  3 &$ 31.03\pm  0.08$ &   30.91$\pm$    0.05	 &   0.06 &   0.00	&   30.88$\pm$0.05 & $30.94 \pm  0.17$  \\
NGC4603 &  5 &$ 33.70\pm  0.09$ &   32.86$\pm$    0.15	 &   0.61 &  -0.11	&         & $33.48 \pm  0.86$  \\
NGC4639 &  4 &$ 31.80\pm  0.08$ &   31.64$\pm$    0.07	 &   0.10 &   0.00	&   31.61$\pm$0.08 & $31.71 \pm  0.21$  \\
NGC4725 &  2 &$ 30.53\pm  0.11$ &   30.44$\pm$    0.06	 &  -0.08 &  -0.05	&   30.38$\pm$0.06 & $30.46 \pm  0.18$  \\
NGC5253 &  6?&$ 27.53\pm  0.14$ &   27.39$\pm$    0.18	 &   0.23 &   0.00	&   27.56$\pm$0.14 & $27.48 \pm  0.34$  \\
NGC5457 &  5 &$ 29.30\pm  0.07$ &   29.23$\pm$    0.07	 &   0.03 &   0.01	&   29.13$\pm$0.11 & $29.26 \pm  0.15$  \\
NGC598  &  5 &$ 24.83\pm  0.12$ &   24.70$\pm$    0.13	 &  -0.28 &  -0.22	&   24.56$\pm$0.10 & $24.77 \pm  0.39$  \\
NGC6822 & 10 &$ 23.38\pm  0.52$ &   23.22$\pm$    0.52	 &   0.00 &   0.00	&         & ($23.30 \pm  0.40$)  \\
NGC7331 &  5 &$ 30.93\pm  0.12$ &   30.80$\pm$    0.11	 &   0.12 &   0.00	&   30.81$\pm$0.09 & $30.86 \pm  0.22$  \\
NGC925  &  6 &$ 29.83\pm  0.06$ &   29.77$\pm$    0.07	 &   0.00 &  -0.05	&   29.80$\pm$0.04 & $29.80 \pm  0.16$  \\
SEXA    & 10 &$ 25.75\pm  0.23$ &   25.78$\pm$    0.26	 &   0.31 &  -0.13	&         & ($25.76 \pm  0.44$)  \\
SEXB    & 10 &$ 26.77\pm  0.18$ &   26.72$\pm$    0.30	 &  -0.18 &  -0.22	&         & ($26.76 \pm  0.35$)  \\
\hline
\end{tabular}
\label{fin}
\end{table*}

\acknowledgements{We thank the HST teams for making their data
available in the literature prior to the end of the project.
We thank R. Garnier, J. Rousseau  and P. Lanoix for having
participated in the maintenance of our Cepheid database.
P.T. acknoledges the support by the Academy of Finland (project
"Cosmology from the local to the deep galaxy universe".
We thank the anonymous referee for valuable remarks.}




\begin{thebibliography}{}
\bibitem{} Alibert Y., Baraffe I., Hauschildt P., Allard F., 1999, A\&A 344, 551
\bibitem{} Alves D.R., Cook K.H., 1995, AJ110, 192 (Alv95)
\bibitem{} Beaulieu J.P., Sasselov D.D., Renault C. et al.: 1997, A\&A 318, L47
\bibitem{} Bono G., Castellani V., Marconi M., 2000, ApJ 529, 293
\bibitem{} Caldwell J.A.R., Coulson I.M., 1987, AJ 93, 1090
\bibitem{} Caldwell J.A.R., Laney C.D., 1991, in Proc. IAU Symp. 148, "The Magellanic Clouds", eds. Haynes R., Milne D., Kluwer, Dordrecht p.249
\bibitem{} Caputo F., Marconi M., Musella I., Santolamazza P., 2000, A\&A 359, 1059
\bibitem{} Cardelli J.A., Clayton G.C., Mathis J.S., 1989, ApJ 345, 245
\bibitem{} Chiosi C., Wood P.R., Capitanio N.: 1993, ApJS 86, 541
\bibitem{} Christian C.A., Schommer R.A., 1987, AJ93, 557  (Chr87)
\bibitem{} Feast M.W., Catchpole R.M., 1997, MNRAS 286, L1
\bibitem{} Ferrarese L. et al., 1996, ApJ 464, 568 (Fer96)
\bibitem{} Ferrarese L. et al., 1998, ApJ 507, 655 (Fer98)
\bibitem{} Freedman W.L. et al, 1994, ApJ 427, 628  (Fre94)
\bibitem{} Freedman W.L., 1988, ApJ 326, 691 (Fr88a)
\bibitem{} Freedman W.L., Madore B.F., 1990, ApJ 365, 186 (Fre90)
\bibitem{} Freedman W.L., Madore B.F., Gibson B.K., et al., 2001, ApJ 553, 47 (HSTKP)
\bibitem{} Freedman W.L., Madore B.F., Hawley S.L. et al., 1992, ApJ 396, 80  (Fre92)
\bibitem{} Freedman W.L., Wilson C.D., Madore B.F., 1991, ApJ 372, 455 (Fre91)
\bibitem{} Gallart C., Aparicio A., Vichez J.M., 1996, AJ112, 1928 (Gal96)
\bibitem{} Gibson, B.K. et al., 1998, astro/ph981003 (unpublished) (Gib98)
\bibitem{} Gibson, B.K. et al., 1999, ApJ 512, 48 (Gib99)
\bibitem{} Gieren W., Fouqu\'e P., Gomez M.: 1998, ApJ 496, 17  (GFG)
\bibitem{} Graham J.A. et al., 1997, ApJ 477, 535 (Gra97)
\bibitem{} Graham, J.A. et al., 1999, ApJ 516, 626 (Gra99)
\bibitem{} Herrnstein J.R., Moran J.M., Greenhill L.J., et al., 1999, Nature 400, 539
\bibitem{} Hughes S.M.G. et al., 1998, ApJ 501, 32 (Hug98)
\bibitem{} Iben I.: 1967, Ann. Rev. Astr. Astrophys. 5, 606
\bibitem{} Kayser S.E. 1967 AJ72, 134         (Kay67)
\bibitem{} Kelson D.D. et al., 1996, ApJ 463, 26 (Kel96)
\bibitem{} Kelson, D. D. et al., 1999, ApJ 514, 614,  (Kel99)
\bibitem{} Kennicut R.C., Stetson P.B., Saha A., et al., 1998, ApJ 498, 181
\bibitem{} Kinman T.D., Mould J.R., Wood P.R., 1987, AJ93, 833  (Kin87)
\bibitem{} Laney C.D., Stobie R.S., 1994, MNRAS 266, 441  (LS)
\bibitem{} Lanoix P., Paturel G., Garnier R., 1999, MNRAS 308, 969
\bibitem{} Lanoix P., Paturel G., Garnier R., et al. 1999a, ApJ. 517, 188
\bibitem{} Lanoix P., Paturel G., Garnier R., et al. 1999b, Astron. Nach. 320, 1
\bibitem{} Lutz T.E., Kelker D.H., 1973, PASP 85, 573
\bibitem{} Macri, L. M. et al., 1999,  ApJ 521, 155 (Mac99)
\bibitem{} Madore B.F., Freedman W.L., 1991, PASP 103, 933
\bibitem{} Maoz, E. Newman J.A.,  Ferrarese L. et al., 1999, Nature 401,  351 (Mao99)
\bibitem{} Mould J.R. et al., 2000, ApJ 528, 655 (Mou99)
\bibitem{} Musella I., Piotto G., Capaccioli M., 1997, AJ114, 976 (Mus98)
\bibitem{} Newman J.A., Ferrarese L., Stetson P.B. et al., 2001, ApJ 553, 562
\bibitem{} Newman, J. A.,  Zepf E.,  Davis M. et al., 1999, ApJ 523, 506 (New99)
\bibitem{} Paturel G., Andernach H., Bottinelli L., et al., 1997, A\&AS 124, 109
\bibitem{} Paturel G., Lanoix P., Garnier R., et al. 1997a, in Proceedings of the ESA Symposium "Hipparcos Venice '97", eds. Perryman, M.A.C. \& Bernacca, P.L.,
\bibitem{} Paturel G., Theureau G., Fouqu\'e P., et al., 2002, A\&A (paper I)
\bibitem{} Perryman M.A.C., Hog E., Kovalevsky J., Lindengren L., Turon C., 1997, in "The Hipparcos and Tycho Catalogues", European Space Agency, SP-1200
\bibitem{} Phelps R.L. et al., 1998, ApJ 500, 763 (Phe98)
\bibitem{} Piotto G., Capaccioli M., Pellegrini C., 1994, AA  287, 371 (Pio94)
\bibitem{} Pont F., Charbonnel C., Lebreton Y., Mayor M., Turon C., 1997, ESA Symp. {\it Hipparcos} - Venice'97, ESA SP-402, ESA Noordwijk, P.699
\bibitem{} Prosser C. F.,  et al., 1999, ApJ 525, 80 (Pro99)
\bibitem{} Rawson D.M. et al., 1997, ApJ 490, 517 (Raw97)
\bibitem{} Saha A., Labhardt L., Schwengeler H. et al., 1994, ApJ 425, 14 (Sah94)
\bibitem{} Saha A., Sandage A., Labhardt L. et al., 1995, ApJ 438, 8 (Sah95)
\bibitem{} Saha A., Sandage A., Labhardt L. et al., 1996, ApJ 466, 55 (Sh96a)
\bibitem{} Saha A., Sandage A., Labhardt L. et al., 1996, ApJS 107, 693 (Sh96c)
\bibitem{} Saha, A. et al, 1999, ApJ 522, 802 (Sah99)
\bibitem{} Saha, A. et al., 1997, ApJ 486, 1 (Sah97)
\bibitem{} Sakai, S. et al., 1999, ApJ 523, 540 (Sak99)
\bibitem{} Sandage A., 1988, PASP 100, 935  (Sa88a)
\bibitem{} Sandage A., Carlson G., 1985b, AJ90, 1019  (Sa85b)
\bibitem{} Sandage A., Carlson G., 1988, AJ96, 1599 (Sa88b)
\bibitem{} Sandage A., Bell R.A., Tripicco M.J., 1999, ApJ 522, 250
\bibitem{} Sasselov D., et al., 1997, A\&A 324, 471
\bibitem{} Silbermann N. A. et al., 1996, ApJ 470, 1 (Sil96)
\bibitem{} Silbermann N. A. et al., 1999, ApJ 515, 1 (Sil98)
\bibitem{} Stetson P.B., 1998, PASP 110, 1448
\bibitem{} Tanvir N. R., Shanks T., Ferguson H. C. et al., 1995, Nature 337, 27 (Tan95)
\bibitem{} Tanvir N., 1997, in "The Extragalactic Distance Scale", Cambridge Univ. Press, Eds. Livio et al.,Cambridge, p91
\bibitem{} Teerikorpi, P. 1987, A\&A, 173, 39
\bibitem{} Teerikorpi, P. , Paturel G., 2002, A\&A 381, L37
\bibitem{} Tully R.B., Fisher J.R., 1977, A\&A 54, 661
\bibitem{} Turner A. et al., 1998, ApJ , 505, 207 (Tur98)
\bibitem{} Udalski A., Szymanski M., Kubiak M., Piertzunski G., Soszynski I., Wozniak P, Zebrun K., 1999, Acta Astronomica 49, 201 (Uda99)
\bibitem{} Walker A.R., 1988, PASP100, 949  (Wal88)
\end{thebibliography}
\end{document}